\documentclass[aps,prd,twocolumn,groupedaddress]{revtex4}
\usepackage{epsfig,amsmath,amssymb}

\newcommand{\dl}{D_L}

\newcommand{\etal}{\eta_{\rm wl}}
\newcommand{\etai}{\eta_{\rm inst}}

\newcommand{\nbar}{\bar{n}}
\newcommand{\dom}{\Delta \Omega}
\def \lleq {\lower0.9ex\hbox{ $\buildrel < \over \sim$} ~}
\def \ggeq {\lower0.9ex\hbox{ $\buildrel > \over \sim$} ~}
\newcommand{\ie}{{\it i.e.} }

\newcommand{\bfn}{\boldsymbol{\rm f}}                                       
                                        
\newcommand{\bh}{\boldsymbol{\rm h}}                                       
\newcommand{\bhT}{\boldsymbol{\rm h}^{\rm T}}                                        
                                        
\newcommand{\C}{\boldsymbol{\rm C}}                                        
                                        
\newcommand{\half}{\frac{1}{2}}  
\newcommand{\beq}{\begin{equation}}
\newcommand{\eeq}{\end{equation}}
\newcommand{\ber}{\begin{eqnarray}}
\newcommand{\eer}{\end{eqnarray}}

\begin{document}

\title{Reconstructing the Properties of Dark Energy using Standard Sirens}

\author{Maryam Arabsalmani}
\email[]{maryam@iucaa.ernet.in}
\affiliation{Inter-University Centre for Astronomy and Astrophysics (IUCAA), 
Ganeshkhind, Pune 411 007, India}
\author{Tarun Deep Saini}
\email[]{tarun@physics.iisc.ernet.in}
\affiliation{Indian Institute of Science, Bangalore, 560 012, India} 
\author{Varun Sahni}
\email[]{varun@iucaa.ernet.in}
\affiliation{Inter-University Centre for Astronomy and Astrophysics (IUCAA), 
Ganeshkhind, Pune 411 007, India}

\date{\today}
\sloppy
\begin{abstract}

Future space-based gravity wave experiments such as the Big Bang
Observatory (BBO), with their excellent projected, one sigma angular
resolution, will measure the luminosity distance to a large number of
gravity wave (GW) sources to high precision, and the redshift of the
single galaxies in the narrow solid angles towards the sources will
provide the redshifts of the gravity wave sources. One sigma BBO beams
contain the actual source only in 68 per cent cases; the beams that do
not contain the source may contain a spurious single galaxy, leading to
misidentification. To increase the probability of the source falling
within the beam, larger beams have to be considered, decreasing the
chances of finding single galaxies in the beams. Saini, Sethi and Sahni
(2010) argued, largely analytically, that identifying even a small
number of GW source galaxies furnishes a rough distance-redshift
relation, which could be used to further resolve sources that have
multiple objects in the angular beam. In this work we further develop
this idea by introducing a {\em self-calibrating} iterative scheme which
works in conjunction with Monte-Carlo simulations to determine the
luminosity distance to GW sources with progressively greater accuracy.
This iterative scheme allows one to determine the equation of state of
dark energy to within an accuracy of a few percent for a gravity wave
experiment possessing a beam width an order of magnitude larger than BBO
(and therefore having a far poorer angular resolution). This is achieved
with no prior information about the nature of dark energy from other
data sets such as SN~Ia, BAO, CMB etc.

\end{abstract}
\maketitle

A remarkable property of our universe is that it is accelerating. The
cause of cosmic acceleration is presently unknown and theorists have
speculated that it might be due to the presence of the cosmological
constant, an all pervasive scalar field called Quintessence, a
Born-Infeld type scalar called the Chaplygin gas, etc. It has also been
suggested that modifications to the gravity sector of the theory, such
as extra dimensional `braneworld' models or $f(R)$ theories, might be
responsible for cosmic acceleration. Establishing the nature and cause
of cosmic acceleration is clearly a paramount objective of modern
cosmology \cite{DE,DE1}. Standard candles in the form of type Ia supernovae
(SNIa) and standard rulers such as baryon acoustic oscillations (BAO)
observed in the clustering of galaxies, have played a key role in
garnering support for the accelerating universe hypothesis. Standard
candles rely on an accurate determination of the luminosity distance to
infer the expansion history and to make a case for cosmic acceleration.
As pointed out in \cite{schutz86,decigo,holz051,arun,sathya} a
complementary probe of the expansion history is available in the form of
gravitational radiation emitted from compact binary objects such as
neutron star - neutron star (NS-NS) binaries, neutron star - black hole
(NS-BH) binaries, or black hole - black hole (BH-BH) binaries.

Indeed, it appears that if the underlying physics behind gravitational
radiation emitted by a NS-NS binary is well understood, then the
luminosity distance to a given redshift, $D_L$, can be established to a
(intrinsic) precision of about 2\% \cite{hirata2010}, making this binary
an excellent {\em standard siren}. However, for a single source the
dominant uncertainty is due to weak lensing, which is 2-3 times this
precision. What is needed additionally, to determine $D_L(z)$, is the
{\em source redshift} of the compact binary emitting gravitational
radiation, and the main systematic uncertainty in this case is the
possible misidentification of the galaxy hosting the binary object (see
e.g. \cite{holz051,arun,sathya}).

\begin{figure}[thb]
\includegraphics[angle=0, width=0.45\textwidth]{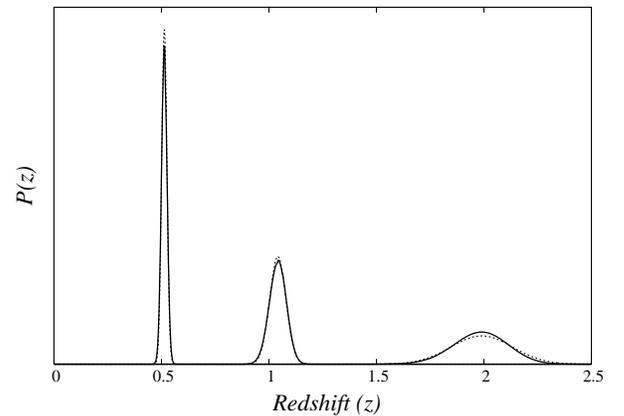}%
\caption{This figure shows the posterior probability distribution $P(z)$
  produced by a linear model for $D_L$ using the Chebyshev polynomials
(\ref{eq:Cheb}).
The solid line shows the {\em exact} expression for $P(z)$ given by (A9)
of S3,
while the dotted line  shows the approximate expression for $P(z)$,
given by (A12) of S3,
and which is used in this paper -- see (\ref{eq:local}).
The two methods of determining $P(z)$ agree very well.}
\label{fig:3p.eps}
\end{figure}

The proposed space-borne gravitational wave observatory LISA \cite{lisa}
was expected to achieve an angular resolution of about $1'$, and the
volume bounded by this angle is expected to contain roughly $30$ objects
at $z \simeq 1$ \cite{holz051}. Its replacement eLISA will have a
considerably lower angular resolution (several degrees) and will
therefore contain far more objects within its field of view
\cite{elisa}. Thus unless the galaxy hosting the binary system can be
unambiguously identified by the electromagnetic afterglow of the merger
event \cite{bense08}, the large number of objects within the eLISA beam
will compound the problem of host identification. Even in
the complete absence of electromagnetic afterglow from GW sources, it
seems possible to obtain useful cosmological information: In a key
paper, MacLeod \& Hogan \cite{macleod} showed that by considering
all the galaxies inside the error box as equally likely sources of a
given GW emission, the Hubble parameter can be measured to an accuracy
of better than a per cent. Their argument is based on observations of
stellar mass black holes inspiralling into massive black holes, the
so-called extreme mass ratio inspiral or EMRI events. These events would
be dominant at $z<1$ where the weak lensing uncertainty is small.
Since galaxies are strongly clustered, and the sources are equally
likely to happen in galaxies irrespective of their clustering, cluster
redshifts would dominate the averaging procedure inside the error box,
thereby providing statistical information about the host redshift.
Our paper builds upon this earlier work \cite{macleod} and shows how one can
iteratively pin down the $D_L$-$z$ relationship even in the absence of
an electromagnetic afterglow from GW sources.

Clearly, then, the major source of uncertainty in determining $D_L(z)$
using standard sirens is caused by the misidentification of the galaxy
hosting the standard siren, and the chances for this to happen increase
with the number of galaxies within the observational beam. Since gravity
wave standard sirens have enormous potential for ascertaining the nature
of dark energy, it would clearly be very desirable to minimize this
source of systematics. Luckily one expects a substantial improvement in
the directional sensitivity of next generation gravitational wave (GW)
experiments. Indeed, space observatories such as DECIGO \cite{decigo},
the Big Bang Observer (BBO) \cite{BBO} and ASTROD \cite{Ni}, currently
in the planning stage, are likely to have a directional sensitivity of a
few arc seconds or better, in which case one might expect only a single
galaxy to fall within the field of view for a large fraction of
observing directions \cite{holz09}. These space based observatories are
expected to measure the equation of state of dark energy to an
unprecedented accuracy \cite{holz09,zhao11}. Although BBO/DECIGO
experiments are in reality at a conceptual stage, for the purpose
of this paper we treat the above mentioned characteristics as a
generic stand in for a future highly advanced gravitational-wave
mission and use the label BBO/DECIGO as a convenient abbreviation 
for such a mission.

Interestingly, even in the
absence of an electromagnetic signature, the source galaxy of the GW
signal can still be singled out from amongst the several galaxies lying
within the observational beam if its redshift is consistent with the
luminosity distance derived from an approximately known cosmology.
Utilizing this idea (\citep{saini10}, hereafter S3) suggested an
iterative scheme to identify the source galaxy of the (unresolved) GW
signal. At the start of the iterative scheme, reliably identified GW
sources --- called `gold plated' (GP) sources, following \cite{holz09}
---give a first estimate of the relationship between the luminosity
distance and redshift (henceforth called the {\em DZ relation}\,$^1$
\footnotetext[1]{Note that the DZ relation should coincide with $D_L(z)$
for idealized measurements.}). The expected number of GP sources depends
crucially on the directional sensitivity of the experiment: good
directional sensitivity will result in a large number of GP sources
whereas the opposite  will be true  for an experiment with poor
sensitivity. The reason for this is simple, an experiment with good
sensitivity will frequently have a single galaxy within its beam and
optical follow ups could establish its redshift.

However, even if one commences with fewer GP sources at the beginning,
one can still improve the DZ relation iteratively as follows. For poor
directional sensitivity (large angular uncertainty) most GW signals
would be unresolved since several galaxies would fall within the (large)
angular beam. However, even in this case, a (rough) DZ relation derived
from GPs can single out one particular galaxy -- the one which is most
consistent with the DZ relation --  to be the source.  This increases
the resolved set, thereby improving the DZ relation, and this procedure
can now be used iteratively. As more and more sources are resolved, the
estimate for the DZ relation improves and eventually saturates at the
point when uncertainty in the redshift of the source is dominated by
instrumental and lensing scatter rather than by our empirical knowledge
of $D_L(z)$.

S3 investigated the efficacy of this method analytically, using the
ensemble average of statistical quantities at each step of the
iteration. Analytically the iteration scheme yields a recursion
relation of the form $N_{j+1} = f(N_{j})$, where $N_j$ is the number
of resolved sources at the $j$th step of the iteration. The limiting
number of resolved sources is then obtained by solving $N = f(N)$,
which is reached after an infinite number of essentially infinitesimal
improvements. In practice, we expect the iterations to freeze much
sooner due to Poisson fluctuations since the number of resolved
sources at each step of the iteration is in reality a rapidly
decreasing random number, and is therefore not expected to change
monotonically.

\begin{figure}[thb]
\includegraphics[width=0.50\textwidth]{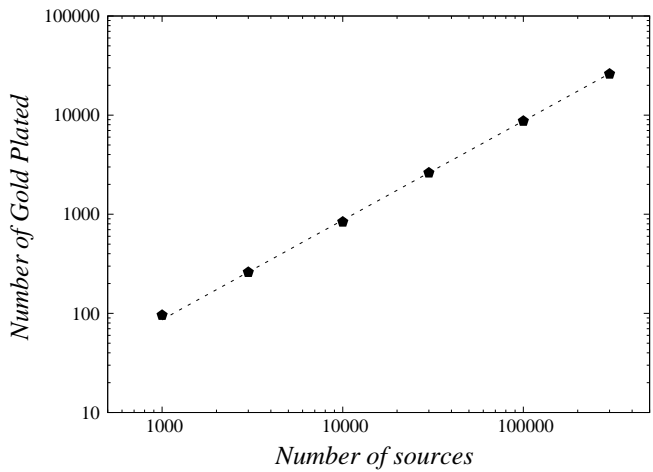}%
\caption{This plot shows the number of gold plated sources
(at the commencement of our iteration) as a function of total number of
events. Note the excellent agreement between the theoretical prediction
in Eq~\ref{eq:NGP} (dashed line) and the simulation (filled hexagons).
}
\label{fig:GP1}
\end{figure}

A crucial aspect of this method is the definition of the \emph{error
box} into which we expect the source to fall. The data only informs us
that the source of the GW signal falls within a solid angle with \emph{a
given probability}; and the measured noisy luminosity distance to the
source, along with the DZ relation, informs us that the source redshift
falls within a redshift interval with a given probability. Therefore,
given an error box we cannot be certain that the true source of the GW
signal lies within it. S3 used one sigma BBO beams with one sigma
redshift range (inferred from noise in the measured luminosity distance)
to define the error box.  However, the assumption of localizing the
source galaxy within the one sigma error box is fraught with
difficulties: if we assume Gaussian noise then the one sigma error box
contains the true source only in 68 per cent of cases, the left-over 32
per cent tail will either contain no galaxy or a non-source galaxy. In
the case when the error box is empty there is no problem since it leads
to an unresolved source, however, if the error box contains the wrong
galaxy, with a redshift different from the true source, the inclusion of
this DZ pair into the GP set will bias the DZ relation. The
misidentified sources are a problem in general, but more seriously,
their specific impact on the iterative process is to drive the inferred
cosmology away from its true value because a biased DZ relation would
deem non-sources as being closer to the biased DZ relation, causing the
number of misidentifications to increase with iterations and thereby
increasing the bias. On the other hand, choosing a larger error box to
decrease the chances of misidentification increases the number of
galaxies falling inside the error box, thereby decreasing the chance of
resolving the GW signal. Clearly, the size of the error box needs to be
chosen in a manner that ensures that misidentifications remain small
but not at the cost of resolving as many sources as possible. A
useful recipe is to start with the smallest error box, implying the
largest number of misidentified sources, and increasing its size until
the bias becomes smaller than the random errors.

On the average, the BBO/DECIGO one sigma angular beam would not contain
the true source in 32 per cent cases but may contain a non-source
galaxy, leading to misidentification. The only way to ensure that beams
do contain the true source galaxy is to consider larger beams and regard
all galaxies falling in, for example, $n\Delta \Omega_{\rm BBO}(z)$ beam
($n \sim \rm few)$, to be potential sources. This ensures that there are
very few beams which do not contain the true source. However, as
mentioned above, this will decrease the total resolved set and therefore
should affect the inferred cosmological constraints. Surprisingly, the
dark energy equation of state can be reconstructed remarkably well even
for $n$ as large as $n=10$, as we demonstrate in this paper.

Our main motivation for the present work is to investigate the above
issues in detail through Monte-Carlo simulation of data; and to explore
choices for the error box that lead to a well determined DZ relation
with minimal bias.

\section{Self-calibration with Monte-Carlo simulated data}

A realistic Monte-Carlo simulation mimicking the outcome of the
BBO/DECIGO experiment requires a careful consideration of the
astrophysical aspects of the problem. First, we have to assume something
about the type of galaxies that would host NS-NS binary systems, and
their abundance, to derive the rate of NS-NS merger events as a function
of redshift. The number of resolved sources as a function of redshift
decides how well the DZ relation is established, which in turn decides
the precision with which one can reconstruct the properties of dark
energy. A detailed prescription for sources of GW signals is presently
not well known and depends on theories of stellar and galactic
evolution, and in this work we continue to use the rates used in Eq~2 of
\cite{holz09}. However, the main issue we are investigating here is the
efficacy of identification of the source galaxies for GW signals using
the DZ relation. Therefore this prescription is entirely adequate for
our purposes. We believe that the achievable precision on cosmology
quoted in our paper is likely to be fairly representative.

For localizing the source galaxy, the prime uncertainty is due to the
presence of other galaxies in the error box; therefore, it is also
necessary to assume something about the spatial distribution of
galaxies.  If the galaxies are clustered in redshift then identification
of the source galaxy becomes relatively more difficult. It is well known
that galaxies cluster in a complex manner, with the clustering depending
both on galaxy type as well as redshift.  Using the two-point
correlation function, S3 gave an estimate of the effect of clustering on
the number of galaxies that are expected to fall within an error box.
Basically, the net effect is to increase the number of galaxies in the
vicinity of the actual source galaxy since the source galaxy is more
likely to lie in a clustered environment. Although it is desirable to
simulate data taking into account the clustering of galaxies, this turns
out to be a difficult task. For the purposes of the present paper we
shall neglect this effect and use the approximation that galaxies are
distributed uniformly randomly at each redshift (our method is described
in greater detail below). As our previous analysis of the two-point
correlation function indicates \cite{saini10}, this will lead to
slightly optimistic estimates of the final achievable cosmological
constraints.

\subsection{Simulating Data}

The simulated data described below has for each GW signal an associated:
redshift of the source galaxy; redshifts of some (or none) non-source
galaxies; the angular beam size at the source redshift; and the noisy
distance estimate along with an estimate for the error in the distance.
We call this data, collectively associated with a single source, as a
\emph{pencil}. We earlier mentioned that if one considers only one sigma
angular resolution then not all BBO beams will contain the source
galaxy. In our simulation each pencil, by construction, contains a
galaxy at the source redshift. Since a one sigma angular beam on an
average contains the source in 68\% of the cases, this amounts to
assuming the total number of simulated sources to be about 50 per cent
higher. Of the 32 per cent beams that do not contain the true source, a
small percentage could contain a non-source galaxy which would
mistakenly be deemed to be a gold plated source. In our prescription
these misidentified sources are not taken into account. As we
demonstrate later, even if we choose an angular beam of size $10\Delta
\Omega_{\rm BBO}$, our method succeeds in resolving the redshifts of
enough GW sources so that the implied constraints on dark energy are
only marginally worse than what one obtains for a one sigma beam. Such
large angular beams are almost certainly going to contain the source
galaxy, therefore this prescription does not lead to any distortion in
our conclusions for larger beam sizes. However, the results quoted for
smaller angular beams would appear to contain smaller bias than would be
the case had we included the misidentified sources.

To populate non-source galaxies in the beams we divide the redshift
range into a large number of redshift bins. Following \cite{holz09}, we
consider sources up to $z=5$. As mentioned earlier, we ignore the
clustering of galaxies, therefore, we assume that galaxies are
distributed uniformly randomly in each redshift bin $\Delta z_{\rm
bin}$. The mean number of galaxies, $\nbar$, lying within the beam at a
given redshift bin depends on the size of the bin $\Delta z_{\rm bin}$
and the angular size of the beam $\dom(z)$, and is given by
\citep{holz09,kaiser,hu}
\begin{equation}
\nbar(z) \simeq \frac{4N_{\Omega}}{h(z) \sqrt \pi} r(z)\,
\exp\left[-r^4(z) \right] \dom(z) \Delta z_{\rm bin}\,,
\label{eq:nbar}
\end{equation}
where we have assumed a small $\Delta z_{\rm bin}$, so the linear
approximation in this equation suffices. Here $r(z) = \int_0^z dz/h(z)$
is the $c/H_0$ normalized coordinate distance, $h(z) = H(z)/H_0$ and
$N_{\Omega} = 1000\,\, \rm arc\, min^{-2}$ is the projected number
density of galaxies consistent with the Hubble Ultra Deep Field
\cite{hudf}.  Given $\nbar$, the probability that there be $k$ galaxies
in the bin is given by
\begin{equation}
\Pr(k) =\nbar^k\exp(-\nbar)/k!\,.
\label{eq:Poisson}
\end{equation}
Using a Poisson random generator we then populate each redshift bin with
non-source galaxies. The one sigma BBO angular resolution,
$\Delta\Omega_{\rm BBO}$, ranges from $1$---$100$ arcsec$^2$ and for
this work we have adopted the redshift dependence of the BBO beam for
NS-NS mergers from Fig~4 of \cite{holz09}. For the BBO beam most of
these bins do not contain any non-source galaxies since $\nbar \ll 1$.
In principle, the redshift distribution of GW sources can differ from
that of galaxies due to the redshift dependent rate of NS-NS mergers,
which we have adopted from Eq~2 of \cite{holz09}. Note that the NS-NS
rate peaks at $z=1$, which is close to where the galaxy density peaks,
so these two distributions are similar to each other (also see figure
\ref{fig:RES_BAD}).

To complete the specification of data we also need a noisy estimate of
the luminosity distance to the GW source. The dimensionless
standard deviation $\sigma_{m}(z)/\dl$ is partly due to the fact that
the luminosity distance to a GW source cannot be measured to better than
about $2\%$ relative accuracy (due to random instrumental noise) and
partly due to weak lensing. The uncertainty due to template fitting
could be, at least, comparable to that due to instrumental error
\cite{cutler2007}, however, in this work we do not include them in our
analysis. The dominant uncertainty is due to lensing. Lensing produces
an asymmetric distribution of magnifications: a majority of the GW
sources are demagnified but a tiny fraction of them are highly
magnified. If the redshifts of the highly magnified sources are resolved
independently then it is possible to handle them statistically; if not
then such sources would either lead to misidentifications or would not
be resolved at all.  However, for our purposes we set aside this
complication$^2$\footnotetext[2]{The asymmetry induced by lensing will
be incorporated in a follow-up work.} and assume that the distribution
is described by a symmetric Gaussian distribution with a dimensionless
standard deviation $\etal(z) = 0.042z$, that is derived from the results
of \cite{holz05}. We add to the lensing standard deviation a fixed
random instrumental/template noise with the dimensionless standard
deviation given by $\etai=0.02$, to obtain the standard error in the
luminosity distance $\sigma_m$ given through \begin{equation}
\frac{\sigma_m}{D_L} = \sqrt{\etal^2+\etai^2}~. \label{eq:sigmam}
\end{equation} Using a Gaussian random number generator with this
standard deviation, we then assign a noisy distance measurement to each
galaxy.

\begin{figure*}[t]
\includegraphics[width=0.50\textwidth]{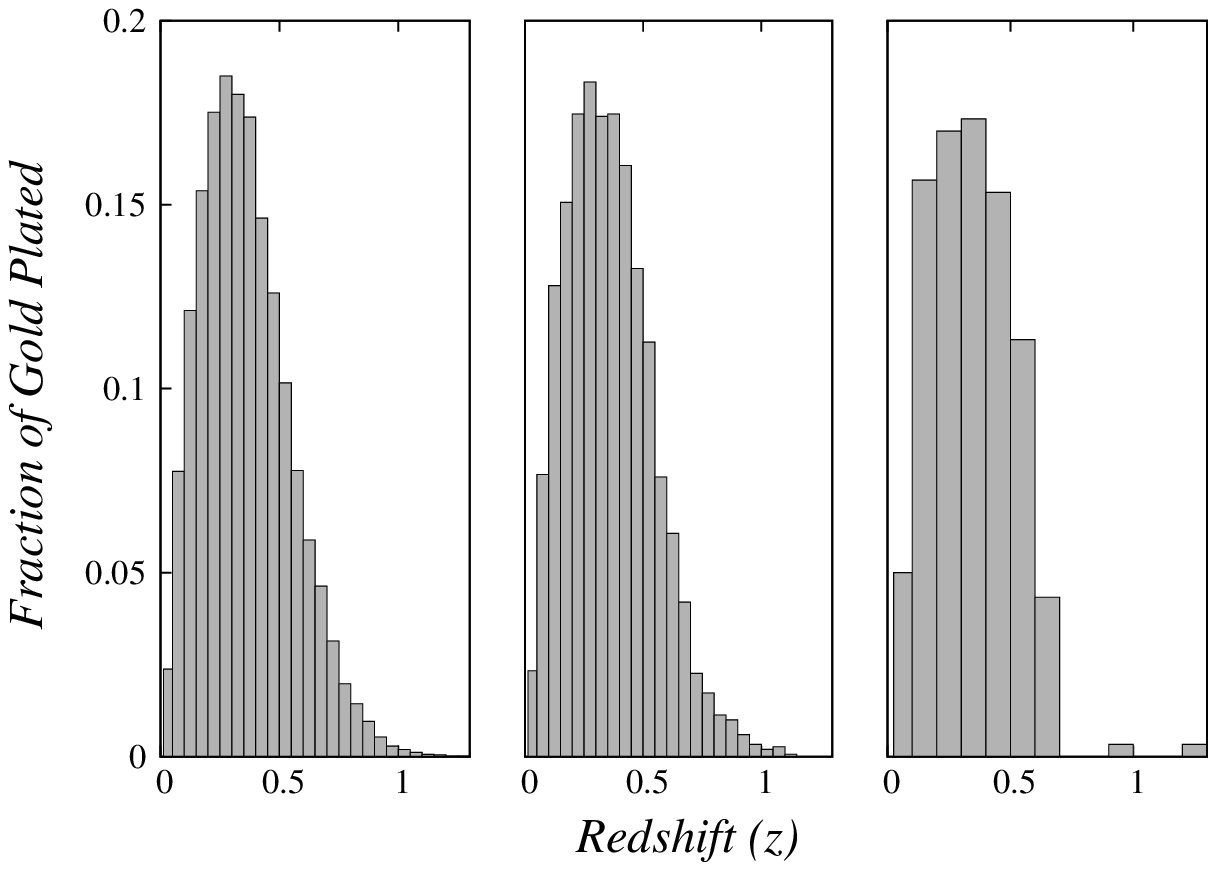}%
\includegraphics[width=0.50\textwidth]{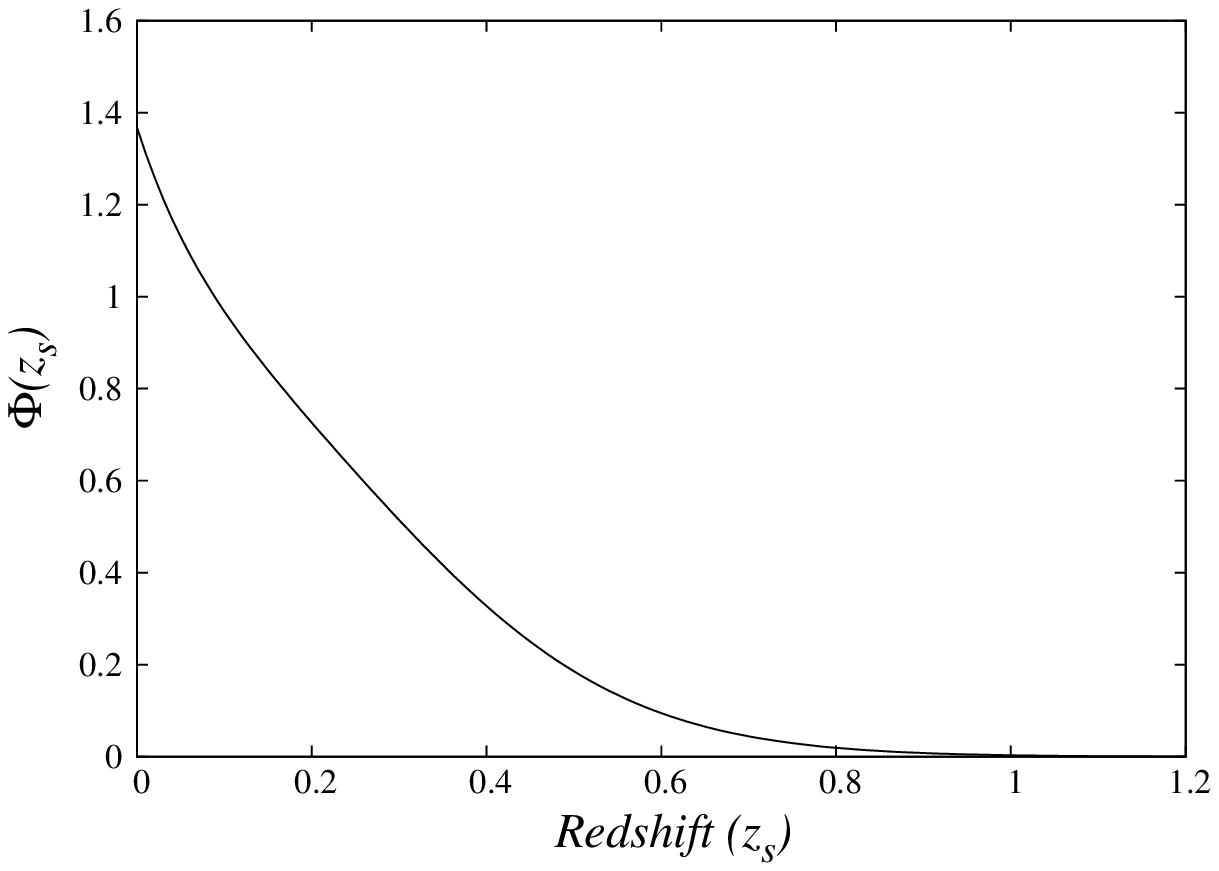}%
\caption{{\it Left panel} displays the distribution of gold plated
sources with redshift, with total number of GW sources being 300,000,
30,000, 3,000, from left to right. {\it Right panel} is a plot of
function $\phi(z_s)$ (Eq.~18). As discussed in the text, these two plots
can be combined to obtain the fractional number of GPs for an $n~
\times$ BBO beam (nBBO) at a given redshift by multiplying the numerical
value in the left panel
and the numerical value in the right panel raised to power $n-1$.}
\label{fig:GPs}
\end{figure*}

\section{Using the DZ relation to increase the number of resolved
sources}

If a pencil contains only a single galaxy (\ie  non-source galaxies are
absent in the beam) then the host galaxy is obviously identified (note
that, by construction, all pencils contain the source galaxy), and so it
can be used to portray the DZ relation at the redshift of the source
(gold plated source). However, as mentioned earlier, the luminosity
distance to a GP source has a non-zero random error due to lensing
scatter and measurement noise.  So using a GP source one can identify
the DZ relation at the redshift of the source only to within a finite
accuracy.

If the number of GPs is comparable to the total number of GW signals,
then we need go no further, but as noted above, this will happen only if
beams at all redshifts are very narrow. The BBO one sigma angular
resolution is sufficiently narrow for this to be the case. However,
since roughly a third of BBO beams will not contain the true source and
might also contain a spurious one -- thereby biasing the DZ relation --
we shall consider beams of larger size$^3$\footnotetext[3]{It could also
happen that the gravity wave observatory that is finally launched would
have a beam width which is larger than that of BBO, in which case the
above considerations would apply.}, to ensure that they more certainly
contain the source galaxy. For larger beams the number of GPs could be
much smaller than the total number of GW signals, in fact, LISA/eLISA
beams are so large that they would have no GPs.

At this stage we have a number of GPs that can be used to further
resolve the pencils that have more than the single source galaxy within
them. To see how this can be done note that the non-source galaxies
would typically have redshifts different from that of the source galaxy.
For a given pencil, the (noisy) DZ relationship inferred
from the GPs provides the probability distribution for the expected
source redshift, $P(z)$, for the signal. Clearly, the galaxy in a pencil
that is closest to the peak of this distribution is more likely to be the
actual source galaxy. The appendix in S3 gives a detailed Bayesian
method for determining this probability distribution using a linear
fitting function (linear in parameters of the model) for the luminosity
distance. According to this method, the GP sources are used to infer the
DZ relation by fitting them to a linear model for the DZ relation,
namely
\beq
\dl(z,\bh) = \sum_i^N h_i f_i(z) = \bhT \bfn\,,
\eeq
where $\bh$ are the $N$ parameters of the model, $f_i$ are $N$
arbitrary functions of redshift, and we have defined $\bfn =
\{f_1(z),f_2(z),\ldots,f_N(z)\}$. The number of terms in the fitting
function is decided by the quality of data: better quality data
requires larger $N$ to adequately fit data. Moreover, the choice of
functional form of $f_i$ should ensure that the fitting function, for
some values of parameters $\bh$, is able to mimic the behavior of the
target DZ relation.

After constructing an appropriate fitting function, the model is fitted
to the resolved sources and the errors on the parameters of the fitting
function are obtained. They are described by the Gaussian distribution
\beq
P(\bh) = \frac{1}{(2\pi)^{N/2} \sqrt{\det{\C}}}\exp\left [-\half
(\bhT-\bhT_0) {\C}^{-1} (\bh-{\bh}_0) \right]\,,
\label{eq:parmcosmo}
\eeq
where ${\C}$ is the covariance matrix and $\bh_0$ are the best fit
parameters. This fit is then used to infer the posterior probability
distribution for the unknown source redshift given the noisy luminosity
distance to the GW signal. The resulting expression is straightforward
but complicated, and is given by Eq.~(A6) of S3.

The Chebyshev polynomials provide a flexible and convenient linear
fitting function for the luminosity distance,
\beq
\dl(z,\bh) = \sum_{I=1}^N h_{I} {\rm ChebI}(z)\,,
\label{eq:Cheb}
\eeq
where ${\rm ChebI}(z)$ is the $I$th order Chebyshev polynomial.
Depending on the quality of data, the order of fit $N$ can be set at a
value where the fit becomes good (in terms of the $\chi^2$ test). The
reason for choosing Chebyshev polynomials over ordinary polynomials is
that the numerical value of these polynomials remains bounded in the
range $(-1,1)$, within the range of the fit. This ensures that the
statistical errors on the coefficients $h_i$ remain similar for all
orders $I$. This is extremely important for translating them to errors
on redshift through $P(z)$, where the covariance matrix needs to be
numerically well behaved for inversion.

We find that this linear model works well most of the time. But in a few
high redshift cases the inferred redshift peak fails to fall reasonably
close to the true source redshift. This happens because noise in the
data affects the fit adversely, especially at high redshifts where the
number of resolved sources is small. The polynomial fit is in some sense
local and does not respect the expectation of monotonicity, therefore it
tries to over-fit any local feature produced by noise; this produces
artifacts that need to be handled individually, making it difficult to
automate the process.

Due to limitations of a linear Chebyshev polynomial based fit, we
considered other alternatives. It is clear that a physics based model
for the luminosity distance does not have these limitations. However, a
crucial unknown is the physics governing the behavior of dark energy,
which is crucial for determining the luminosity distance. The unknown
physics of dark energy is usually encapsulated in the form of a fluid
model for dark energy, with an equation of state, $p=w\rho$. The unknown
function $w$ is then thought of as a function of redshift $w(z)$, and
can be parameterized in terms of a suitably versatile fitting
function$^4$ \footnotetext[4]{In other approaches it is also possible to
work with a fitting function for the Hubble parameter $H(z)$ as
discussed in \cite{DE1}.}. For our work we follow the following
reconstruction procedure which incorporates the CPL fitting function
\cite{CPL} for $w(z)$ given below:
\begin{equation}
\frac{D_L(z)}{1+z} = \frac{c}{H_0} \int_1^{1+z} \frac{dx}{H(x)}
\label{eq:DL}
\end{equation}
where,
\ber
H^2(z) &=& H_0^2 \lbrack \Omega_M (1+z)^3 + \Omega_{\rm DE}\rbrack^2~,\nonumber\\
\nonumber\\
\Omega_{\rm DE} &=& (1-\Omega_M) \exp{\left\lbrace 3 \int_0^{z}
\frac{1+w(z)}{1+z} dz\right\rbrace }~,\nonumber\\
w(z) &=& p_{\rm DE}/\rho_{\rm DE} = w_0 + \frac{w_1z}{1+z}~,
\label{eq:EOS}
\eer
with $\Lambda$CDM corresponding to $w_0=-1,\,w_1=0$.
The CPL ansatz 
produces reasonable fits 
if dark energy has a slowly varying equation
of state. 

However, from the perspective of the iterative scheme which we develop
in this paper, this fitting function has a problem: it depends
non-linearly on the model parameters $\Omega_M$, $w_0$ and $w_1$. The
posterior probability $P(z)$ for the source redshift can be analytically
obtained only if the luminosity distance depends linearly on its
parameters (see appendix of S3). A possible remedy is to:
(i) use the fitting function (\ref{eq:DL}), (\ref{eq:EOS}) to the
simulated data to obtain the best fit parameters ($\Omega_{M0},
w_{i0}$); (ii) then linearize (\ref{eq:DL}) in $\Omega_M$ and $w_i$
through a Taylor expansion about the best fit parameters:
\begin{equation}
D_{\rm Linear}(z,\theta_i) \equiv D_L(z; \theta_{i0}) + \sum_i^3
(\theta_i-\theta_{i0}) \frac{\partial D_L(z; \theta_i)}{\partial
\theta_i} \,
\end{equation}
where, for brevity, $\{\theta_i\} \equiv \{\Omega_M, w_0, w_1 \}$; the
derivatives are evaluated at the best fit parameters $\theta_{i0}$.
This function is linear in parameters $\theta_i$ and can be used
to analytically marginalize over the model parameters to obtain $P(z)$.
If the original fit is tight, in the sense that the errors on
parameters $\theta_i$ are small, then the linearized fitting function
$D_{\rm Linear}(z,\theta_i)$ serves as a close approximation to the
original.

It turns out that a further simplification is possible that makes our
task much simpler. The posterior probability distribution given in
Eq.~A9 of S3 is not a simple Gaussian distribution. The distribution
has a peak at the redshift that the best fit model predicts for the
measured noisy distance to the GW signal but its redshift dependence can
be complicated. It is however possible to systematically extract its
leading \emph{local} behavior to obtain the Gaussian distribution
\beq
P(z) = \frac{1}{\sqrt{2\pi} \sigma_z}
\exp\left[-\frac{(z-z_0)^2}{2\sigma_z^2} \right]\,,
\label{eq:local}
\eeq
where $\sigma_z= \sqrt{(\sigma_m^2 + \sigma_c^2)}/D'_L$; the standard
error $\sigma_m$ is from Eq.~\ref{eq:sigmam}, and $\sigma_c$ is the
cosmology error defined in the last section of the Appendix in S3; $D'_L
= \partial D_L(z,\theta_{i0})/\partial z|_{z=z_0}$; the parameter $z_0$
is the best fit redshift inferred for the GW signal and is computed from
the prescription given there, but basically it is the
redshift inferred for the noisy distance estimate via the best fit model
$D_L(z;\theta_{i0})$. Note that due to noise in the
measured luminosity distance, $z_0$ need not coincide with the true
source redshift.

In Fig~\ref{fig:3p.eps} we plot a comparison of $P(z)$ produced by the
local Gaussian approximation (\ref{eq:local}) and the more exact
expression A9 of S3 (both distributions are produced by using a
Chebyshev polynomial based fit). We find that the Gaussian approximation
agrees very well at all redshifts. In
fact, the agreement becomes better with an increase in the number of
resolved sources because the error bars on parameters become smaller,
and A9 of S3 starts approaching a Gaussian distribution. Therefore,
the local Gaussian approximation is adequate for our purposes.

\begin{figure*}[t]
\includegraphics[width=0.5\textwidth]{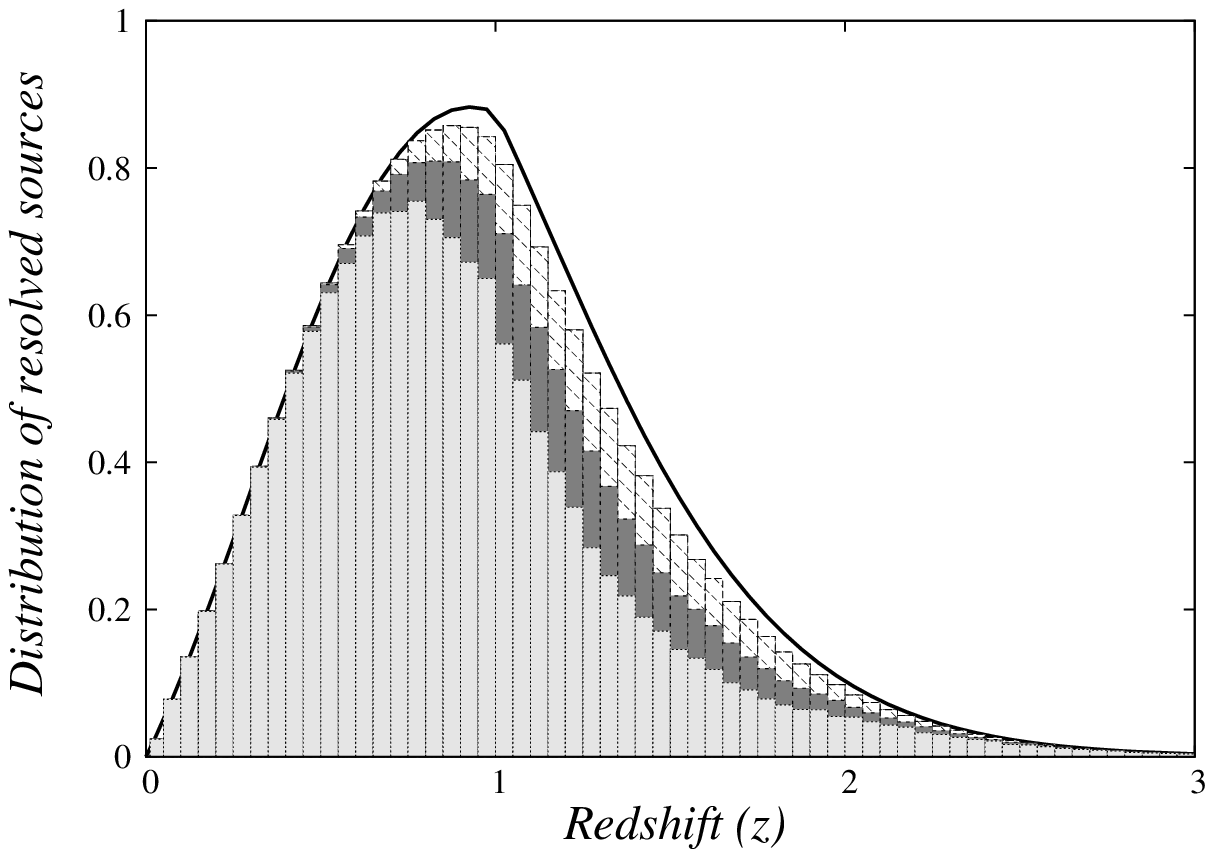}%
\includegraphics[width=0.5\textwidth]{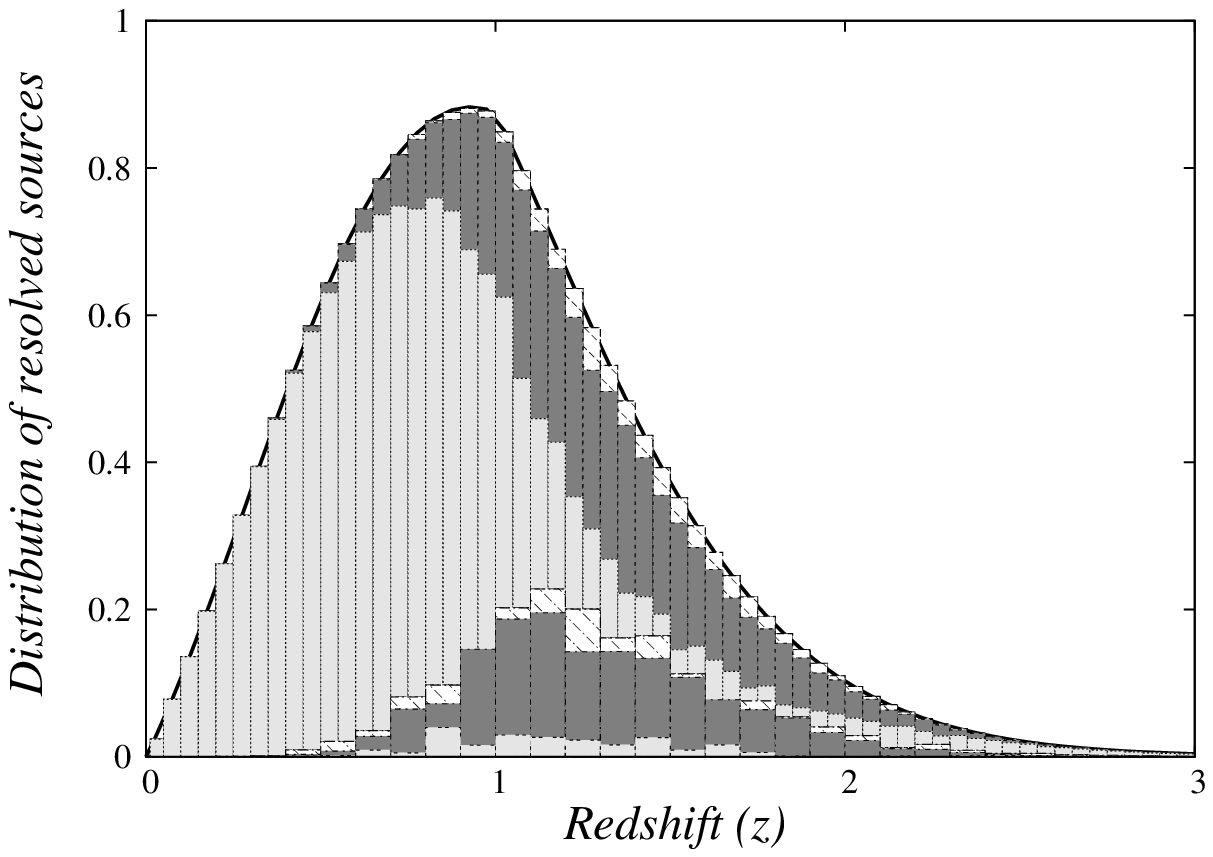}%
\caption{{\it Left panel} shows the normalized distribution of {\em resolved}
GW sources at the end of the iteration. The solid line is the redshift
distribution of all sources assuming BBO resolution; dashed lines show
the redshift distribution of {\em resolved} sources obtained after
assuming that the source lies within
$1\sigma$ of the peak in the probability distribution $P(z)$ defined in
(\ref{eq:local}).
In other words the dashed histogram corresponds to $n=1$ in
(\ref{eq:BBO})
and $m=1$ in (\ref{eq:redshift_range}). Dark gray shows the same but for
a $2\sigma$ allowed region for sources, corresponding to $m=2$ in
(\ref{eq:redshift_range}).
Finally the light gray shade reflects a
$3\sigma$ allowed region for sources and corresponds to
 $m=3$ in (\ref{eq:redshift_range}).
Note that the number of resolved GW sources is comparable to the
total number of GW sources for $m=1$ and is only slightly worse for
$m=2,3$. {\it Right panel} discusses the issue of
source misidentification. Clearly by restricting GW sources to lie
within
an $m\sigma$ region of the peak
of the probability distribution $P(z)$ we open up the possibility of
source misidentification which is larger for smaller $m$.
This is clearly revealed in the inset to the right panel (the shading
scheme
is the same as in the left panel). We see from this panel that fewer
sources are misidentified
for $3\sigma$ (light gray) than for $1\sigma$ (dashed).
The main panel on the right shows the distribution of resolved sources
{\em after}
misidentified sources have been identified. Thereafter misidentified
sources do not play
any role in the iterative scheme which, in the case of the dashed
region, coincides with
the solid line corresponding to the redshift
distribution of all GW sources.
}
 \label{fig:RES_BAD}
\end{figure*}

\subsection{ Error Box}

The local approximation for $P(z)$ in (\ref{eq:local}) is convenient in
that it enables us to define the error box for the source location in
terms of the variance $\sigma_z$ of the local approximation
Eq~\ref{eq:local}. Note that this task would be considerably more
complicated had we used A9 of S3, which is not described by a
few simple parameters as the Gaussian distribution is. To define our
error box, let us first choose the beam size in which we expect the
source to lie as 
\beq
\Delta\Omega(z)=n \Delta \Omega_{\rm BBO},
\label{eq:BBO}
\eeq
 where $\Delta \Omega_{\rm BBO}$ is
the redshift dependent one sigma resolution of the BBO experiment. The
beam has been chosen as a multiple of the BBO resolution to ensure a
high probability for the source to fall inside the beam.
(This also accommodates the possibility that, as in the case of LISA
$\to$ eLISA, the space mission which finally flies has a larger beam
width than BBO.) Similarly, we can define the redshift range in which
the source is likely to lie as
\beq
\Delta z = 2m\sigma_z, 
\label{eq:redshift_range}
\eeq
which is centered at the peak of the probability distribution $P(z)$.
 For simplicity, the redshift range has been chosen
to be a multiple of the one sigma
range obtained from $P(z)$.

\subsection{Iterative Scheme}

We generate pencils with different values of $n$ in (\ref{eq:BBO}). As
mentioned above, larger values of $n$ increase the chances of the true
source of the gravity wave signal lying within the beam, but have the
obvious side effect that the number of non-sources in the beam also goes
up. Clearly, with a larger number of non-sources in each beam, the
number of gold plated sources (GPs) in the set goes down,
therefore, $n$ cannot be chosen too large, otherwise there will be no
GPs with which to start our iterative process. The GPs
are used to derive an initial estimate of the luminosity distance
which is used to infer $P(z)$. Knowing $P(z)$ we repeat the process of
going through each pencil and checking
whether any (single) galaxy is present in the redshift range $
z_0-m\sigma_z < z < z_0+m\sigma_z$, where $z_0$ corresponds to the peak
of $P(z)$. Galaxies inside pencils where this test succeeds are referred
to as {\em resolved galaxies} and added to the burgeoning inventory of
sources. This process is repeated until there is no
further increase in the number of resolved sources and the iterative
method saturates.
Now the final value of the luminosity distance is used to determine
$w(z)$ -- the equation of state of dark energy, using (\ref{eq:DL}) and
(\ref{eq:EOS}). Our self-calibrating iterative scheme is summarized below (convergence
being reached after $N$ steps):
\ber
&{\cal N}^{(1)}_{GP} ~~\longrightarrow~~ D_L^{(1)}(z) ~~\longrightarrow~~ P^{(1)}(z)\nonumber\\
P^{(1)}(z) ~~\Longrightarrow~~ &{\cal N}^{(2)}_{GP} ~~\longrightarrow~~ D_L^{(2)}(z)
~~\longrightarrow~~ P^{(2)}(z)\nonumber\\
&\cdots\cdots\cdots\cdots\nonumber\\
P^{(N-1)}(z) ~~\Longrightarrow~~ &{\cal N}^{(N)}_{GP} ~~\longrightarrow~~ D_L^{(N)}(z)  
~~\longrightarrow~~ {\bf w(z)}~. \nonumber\\
\label{eq:iterative}
\eer
Note that as more resolved sources are added, the
shape of $P(z)$ becomes peakier and narrower (smaller $\sigma_z$), which
helps in
picking new resolved sources from amongst contender galaxies. This
practice, outlined in (\ref{eq:iterative}), is continued until
convergence is reached and no new source galaxies can be added to the
resolved sample.

Let us illustrate this with an example. Supposing of the 1000 pencil
beams that have been generated only 30 have a single galaxy falling
within them and the remaining beams contain two or more galaxies.
Then we can safely assume that these 30 galaxies host GW sources and,
using
optical/IR observations, determine their redshifts. This establishes for
us our initial DZ relationship, namely $D_L^{(1)}(z)$ in
(\ref{eq:iterative}), where ${\cal N}^{(1)}_{GP} = 30$. Next, knowing
$D_L^{(1)}(z)$ we determine the probability distribution function,
$P^{(1)}(z) \equiv P^{(1)}(z|D_L^{(1)})$, through (\ref{eq:local}).
Clearly our knowledge of $P(z)$ will inform us which of the two or more
candidate galaxies (in each of the additional 970 beams) is a
potential gravity wave source, since this galaxy would lie
closer to the peak of $P(z)$ than its companions. Including the resolved
galaxy, and others like it, will establish an enlarged sample, and this
procedure will be followed iteratively (for $N$ steps)
until no new resolved galaxies can be added, at which point we say that
convergence has been reached and the final value of $w(z)$ is determined
from $D_L^{(N)}(z)$ using (\ref{eq:DL}) \& (\ref{eq:EOS}).
If the error box is small (small $n$ and $m$) then it follows that the
chances of resolving a GW signal increases since it is more likely that
we will find only a single galaxy within the error box. However, a very small error
box would also imply that the probability of the true source galaxy lying within 
it be small, leading to a larger number of misidentifications 
in this case. Misidentifications have the pernicious effect of biasing the DZ
relation away from its true value, resulting in a positive feedback on
the chances of misidentifications during later iterations and 
converting the biasing of the DZ relation into a runaway process.  
In the other extreme case when the
error box is very large, the possibility of misidentification goes
down but the chances of resolving GW signals goes down as well, resulting in a 
decrease in the quality of constraints obtained for cosmology. Clearly then,
an optimum
value for the error box should be such that the constraints on cosmology
are tightest while at the same time the bias is below the statistical scatter. 

\section{Results}

At commencement, a small number of gold plated (GP) sources is required in order to
obtain the zeroth order cosmology to start our iterative process of
statistical resolution of GW sources. The probabilistic expectation
value of the number of GP sources for a given number of total sources
and the beam size can be estimated as follows: Dividing the redshift
range into a large number of bins, the mean number of galaxies
$\nbar(z_i)$ in a redshift bin can be estimated through Eq~\ref{eq:nbar}
using the angular beam $\Delta \Omega (z_s)$, where $z_s$ is the
redshift of the source galaxy; and $\Delta z_{\rm bin}$, where $z_i$ is
the redshift of the bin.

It is worth noting that the probability for a 	
source to be a GP depends on the angular resolution with which the
source can be resolved, therefore the angular resolution of the beam is
evaluated at $z_s$, thereby making $\nbar(z_i)$ a function of
$z_s$, which we notate explicitly below. As mentioned earlier, the
BBO angular resolution which we have used is taken from Fig~4 of
\cite{holz09}, corresponding to NS-NS mergers. Our prescription puts the
actual source galaxy in each pencil, therefore, the probability that the
pencil contains no other galaxy can be obtained from the probability
that the redshift bin at $z_i$ is empty, which is $\Pr(0) =
\exp[-\nbar(z_i; z_s)]$ (see Eq~\ref{eq:Poisson}). Clearly, a pencil
will give a GP source if all
bins are empty$^5$\footnotetext[5]{Recall that according to our
prescription an empty pencil beam is redefined so as to contain the
source galaxy.} giving a probability
\begin{equation}
\Pr(z_s; {\rm Gold\,Plated} )= \exp\left[-\sum_{z_i} \nbar(z_i;
z_s)\right]~.
\end{equation}
In the expression for $\nbar(z_i; z_s)$ (Eq~\ref{eq:nbar}), we see
that it is proportional to the bin size $\Delta z_{\rm bin}$. The
probability $\Pr(z_s; {\rm Gold\,Plated} )$, however, is understood
to be obtained in the limit of infinitesimal bin size, in which case
the sum in the above expression will be replaced with an integration.

For a small enough bin size, defining $f_s$ as the fraction of GW
sources at redshift $z_s$, the fraction of gold plated GW sources at
$z_s$ is given by
\begin{equation}
f_{\rm GP}(z_s) = f_s(z_s) \exp\left[-\sum_{z_i} \nbar(z_i;z_s)\right]
\end{equation}
and the total number of GPs by
\begin{equation}
N_{\rm GP} = N_{\rm Total} \sum_{z_s}
f_s(z_s)\exp\left[-\sum_{z_i} \nbar(z_i;z_s)\right]
\label{eq:NGP}
\end{equation}

Fig~\ref{fig:GP1} shows the number of gold plated gravity wave sources
as a function of total number of GW sources obtained for the BBO angular
beam. As predicted by (\ref{eq:NGP}), the theoretical expectation that
$N_{\rm GP} \propto N_{\rm Total}$ is found to be true in our
simulations. We find that even for a small total number of sources
($\sim 1000$) the GP set is large enough ($\sim 100$). The left panel of
Fig~\ref{fig:GPs} plots the function $f_{\rm GP}^{\rm BBO}(z_s)$. We
find that the distribution is independent of the total number of
sources, in accordance with theoretical expectation. It is important to
note that {\em the distribution of sources in redshift is wide enough
for obtaining a good starting DZ $\equiv D_L(z)$ relation}. Note also
that the peak of the redshift distribution $f_{\rm GP}$ of GPs (at $z
\sim 0.3$) does not coincide with the peak of the galaxy distribution
($z\sim 1$). This is due to the fact that there is an additional
redshift dependence due to the redshift dependent beam size
\cite{holz09}. The BBO angular resolution is better at smaller redshift;
however the number of sources first increase, peak at $z\sim 1$, and
then decrease with the redshift; the net effect therefore is to shift
the peak of the GPs redshift distribution to a lower redshift.

For larger (than BBO) angular resolution, the number of GPs decreases
since the beam, in many cases, becomes too large to accommodate
only a single galaxy. This effect can be estimated in terms of the
function $f^{\rm nBBO}_{\rm
GP}(z)$ for
$\Delta
\Omega(z) = n\Delta \Omega_{\rm BBO}(z)$, which can be easily shown to
be given by
\begin{eqnarray}
\nonumber
f^{\rm nBBO}_{\rm GP}(z_s) &=& f_s(z_s) \exp\left[-n\sum_{z_i}
\nbar_{\rm
BBO}(z_i;z_s)\right]\\
&=& f_{\rm GP}^{\rm BBO}(z_s) \phi^{n-1}(z_s)\\
\phi(z_s)&\equiv & \exp\left[-\sum_{z_i} \nbar_{\rm
BBO}(z_i;z_s)\right]\,.
\end{eqnarray}
We find that at each redshift, the number of GPs predicted for the
$n\Delta \Omega_{\rm BBO}(z)$ beam is smaller than that for $\Delta
\Omega_{\rm BBO}(z)$ beam by a redshift dependent multiplicative factor
$\phi^{n-1}(z_s)$. The right panel of Fig~\ref{fig:GPs} plots the
function $\phi(z)$. We see that for $z>0.1$, $\phi < 1$ is a small
fraction; therefore this factor decreases very rapidly with $n$. The
multiplicative factor increases with $n$ for $z<0.1$ but since the
fraction of sources is small at small redshift, this does not cause an
abnormal increase in GPs at low redshift with increasing $n$. Although
mathematically speaking this factor blows for large values of $n$, at
a large enough $n$ the Poisson statistic will cease to hold, making this
argument invalid.

We simulated $300,000$ pencils corresponding to three years of
cumulative BBO data. The resolved samples were fitted, at various stages
of iteration, with the non-linear model for $D_L(z)$ described by
(\ref{eq:DL}) in which the CPL equation of state (\ref{eq:EOS}) has been
used. The model was linearized over the polynomial coefficients and the
local approximation discussed earlier was then used to derive $P(z)$.
The fiducial model chosen for all our
simulations was a flat $\Lambda$CDM model with $\Omega_m=0.3$. In all cases
(described below) the iterations freeze out after about ten steps
implying that convergence had been reached, \ie $N=10$ in
(\ref{eq:iterative}). 

In the left panel of Fig~\ref{fig:RES_BAD} we plot the total number of
resolved GW sources obtained at the end of our iterative run with $n=1$ in
(\ref{eq:BBO}) and $m=1,2,3$ in (\ref{eq:redshift_range}). It is
interesting that the number of resolved GW sources is comparable to the
total number of GW sources for $m=1$ and is only slightly worse for
$m=2,3$. The inset in the right panel shows the distribution of
misidentified sources (the redshift has been identified incorrectly).
Recall that $n=1$ corresponds to the $1\sigma$ angular resolution for
BBO and, as pointed our earlier, this implies that roughly a third of
all beams will not contain {\em any GW sources at all ~}! However, in
our simulations we ensure that all pencils do contain the true source,
therefore figure \ref{fig:RES_BAD} does not include the effect of
misidentification of sources due to small beam size. The other important
cause for misidentifications is due to a choice of too small an allowed
redshift range, which is what the inset in figure \ref{fig:RES_BAD}
illustrates (the number of misidentified sources increases as $m$
decreases). The fraction of misidentified sources peaks at $z\sim 1$
because, as mentioned earlier, the BBO beam monotonically becomes wider
at larger redshift and so has a larger number of non-source galaxies
falling into it; therefore the expected peak redshift for
misidentifications is expected to be larger than the peak redshift of
the galaxy distribution (albeit only slightly).

\begin{figure}[t]
\includegraphics[width=0.45\textwidth]{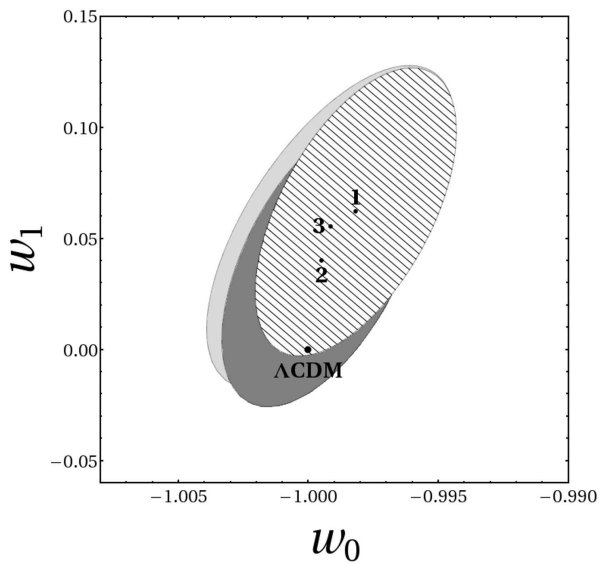}%
\caption{ Constraints on $w_0$ and $w_1$ with misidentified sources
  included in the resolved sample. The beam size is fixed at the one
sigma angular resolution of BBO, corresponding to $n=1$ in
(\ref{eq:BBO}), while the redshift range in which the source is expected
to lie is given by (\ref{eq:redshift_range}), namely $\Delta z =
2m\sigma_z$. The lined region corresponds to
  $m=1$,  dark gray to $m=2$, and light gray to $m=3$ in (\ref{eq:redshift_range}).
In all three cases the fiducial $\Lambda$CDM model
($w_0 = -1, w_1 = 0$) is included within the one sigma contour, but the fit improves for
$m=2,3$, corresponding to a decrease in uncertainty of the source redshift.
}
\label{fig:1.eps}
\end{figure}

In Fig~\ref{fig:1.eps} we plot the constraints on $w_0$ and $w_1$
obtained with $n=1$ and $m=1,2,3$. This figure includes the
misidentified sources in order to illustrate how misidentified sources
can bias the cosmology. Although
the centers of the three one sigma regions (corresponding to three
different values of $m$) predict a non-zero variation in dark energy
(non-zero $w_1$), the fiducial $\Lambda$CDM model
($w_0 = -1, w_1 = 0$) does fall within the one
sigma contour. Also note that larger values of $m$, corresponding to
fewer misidentifications, are more consistent with the fiducial $\Lambda$CDM model. 
Note also that although the 1$\sigma$ contour increases slightly for larger $m$,
this effect is pretty marginal.

\begin{figure}[t]
\includegraphics[width=0.45\textwidth]{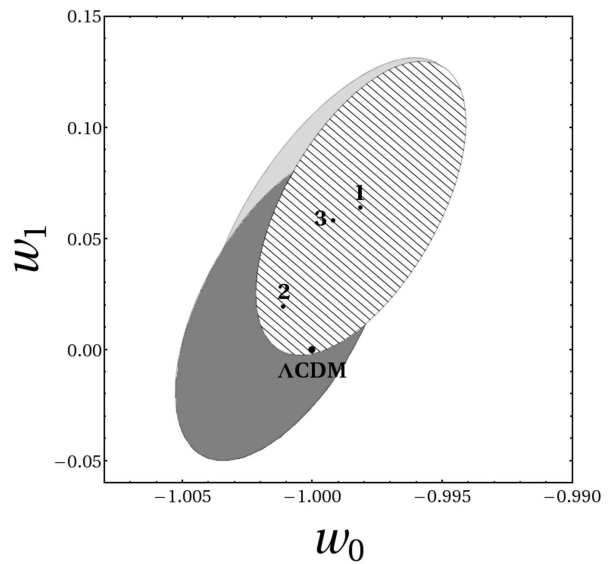}%
\caption{Same as figure \ref{fig:1.eps} but after misidentifications
have been removed in the resolved sample. The beam size is fixed at
the one sigma angular resolution of BBO, corresponding to $n=1$ in
(\ref{eq:BBO}). The lined region corresponds to $m=1$,  dark gray to
$m=2$, and light gray to $m=3$ in (\ref{eq:redshift_range}). In all
three cases the fiducial $\Lambda$CDM model ($w_0 = -1, w_1 = 0$) is
included within the one sigma contour. Note that for $m=2$ (dark gray)
the fit has improved substantially in comparison with figure
\ref{fig:1.eps}.
}
\label{fig:2.eps}
\end{figure}

\begin{figure}[t]
\includegraphics[width=0.45\textwidth]{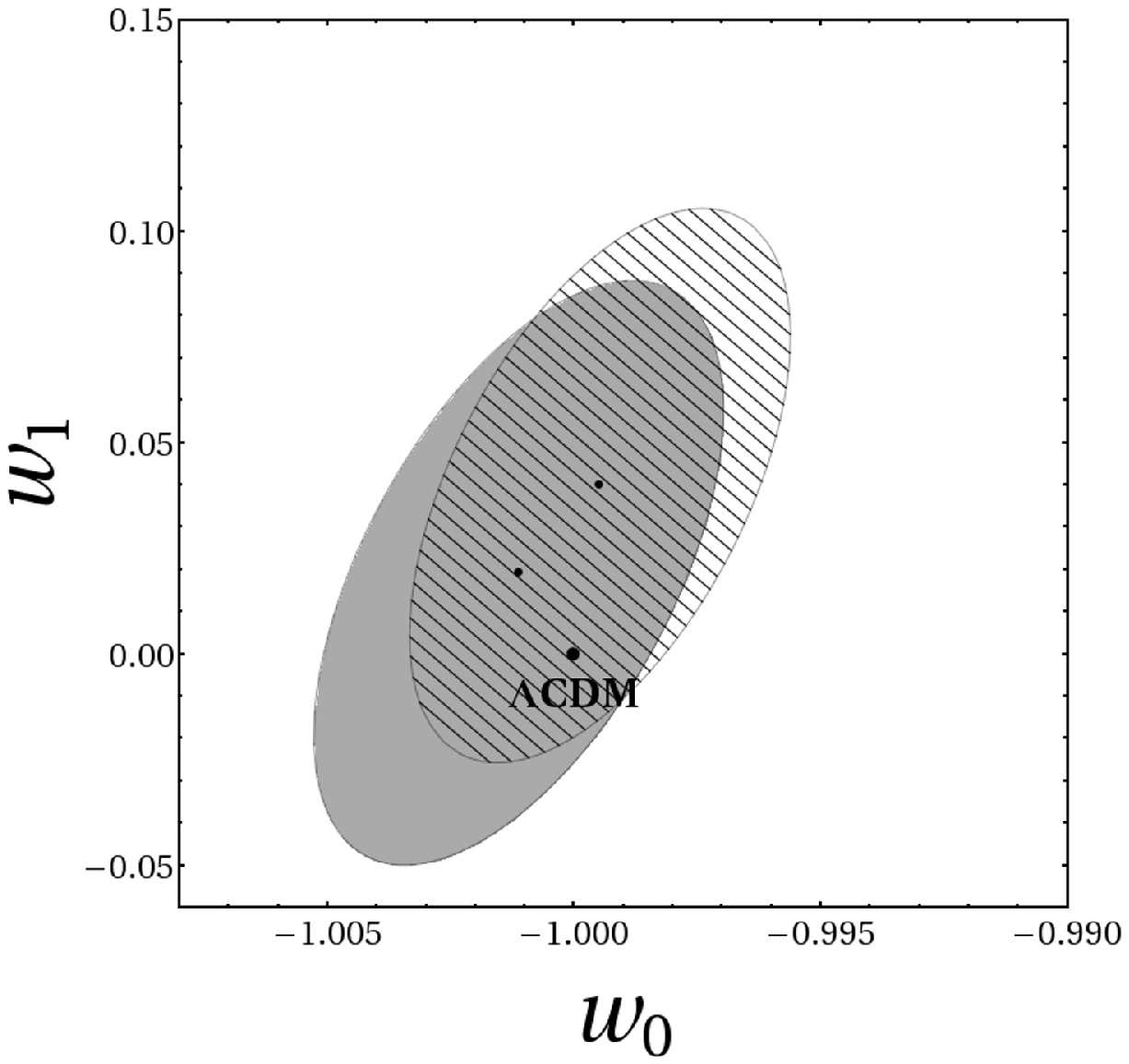}%
\caption{
This figure shows the effect of including/excluding misidentified GW
sources. The beam size is kept at the BBO one sigma angular resolution
corresponding to $n=1$ in (\ref{eq:BBO}) and the redshift uncertainty of
the source is two sigma: $m=2$ in (\ref{eq:redshift_range}). In this
figure the dashed region shows the fit with misidentified sources
included and the gray region after misidentified sources have been
excluded. As expected, the  one sigma contour is in better agreement
with the fiducial $\Lambda$CDM model ($w_0 = -1, w_1 = 0$) if
misidentified sources have been excluded from the sample.
}
\label{fig:4.eps}
\end{figure}

This may not be as surprising as it appears. The BBO beam is so narrow
that the probability of finding a galaxy within it is small. The one
sigma redshift range corresponding to $m=1$ would contain the source
galaxy in roughly 68 per cent pencils, while in the remaining cases most
pencils would not contain any galaxy at all. Therefore, the total number
of misidentifications would be small, and their main role would be in
biasing the DZ relation and not so much in controlling the tightness of
fit. This is illustrated in Fig~\ref{fig:2.eps}, which is identical to
Fig~\ref{fig:1.eps} in all respects other than the fact that here we
have removed the misidentified sources. The agreement with the fiducial
model is slightly better in this case (especially for $m=2$). The
scatter in best fit value is due to the slightly different number of
resolved sources in the three cases, however, all three regions are
consistent with each other. Fig~\ref{fig:4.eps} isolates the effect of
including or excluding the misidentified sources. 
It is clear from this figure that biasing, due to choosing too narrow a
redshift range for the source redshift, is present but is smaller than
the statistical errors on cosmological parameters.

\begin{figure}[t]
\includegraphics[width=0.45\textwidth]{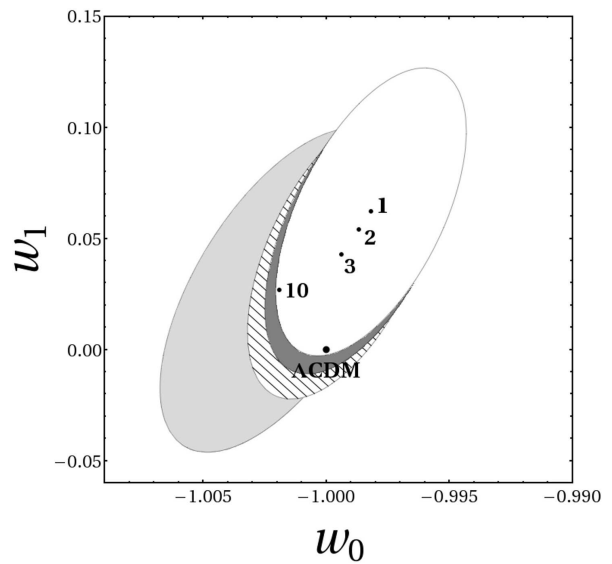}%
\caption{In this figure we illustrate the effect of increasing the beam
width on the reconstruction of the equation of state of dark energy. The
uncertainty in source redshift remains fixed at $m=1$ in
(\ref{eq:redshift_range}) while the beam width increases as $n=1,2,3,10$
in (\ref{eq:BBO}), which corresponds to $1\sigma, 2\sigma, 3\sigma$ and
$10\sigma$ times the BBO value. White corresponds to $n=1$, dark gray to
$n=2$, dashed to $n=3$ and light gray to $n=10$. Increasing the beam
size decreases the overall number of resolved GW sources using which
$w(z)$ is reconstructed, and therefore slightly increases the area of
the one sigma contour. It is interesting that the fiducial $\Lambda$CDM
model ($w_0=-1, w_1=0$) lies within the $1\sigma$ contour for all values
of $n$. }
\label{fig:7.eps}
\end{figure}

\begin{figure}[t]
\includegraphics[width=0.45\textwidth]{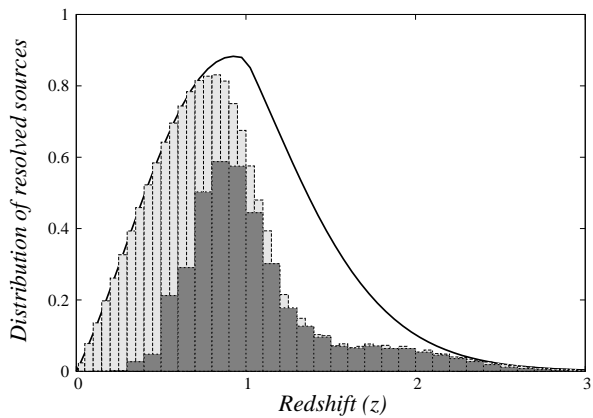}%
\caption{The normalized distribution of resolved
sources at the end of our iteration. The solid line is the redshift
distribution of all sources. The total fraction of resolved (light gray)
and misidentified (dark gray) sources is plotted as a
function of redshift for $n=10$ in (\ref{eq:BBO}) and $m=1$ in (\ref{eq:redshift_range}).}
 \label{fig:10BBO}
\end{figure}

\begin{figure*}[t]
\includegraphics[width=0.50\textwidth]{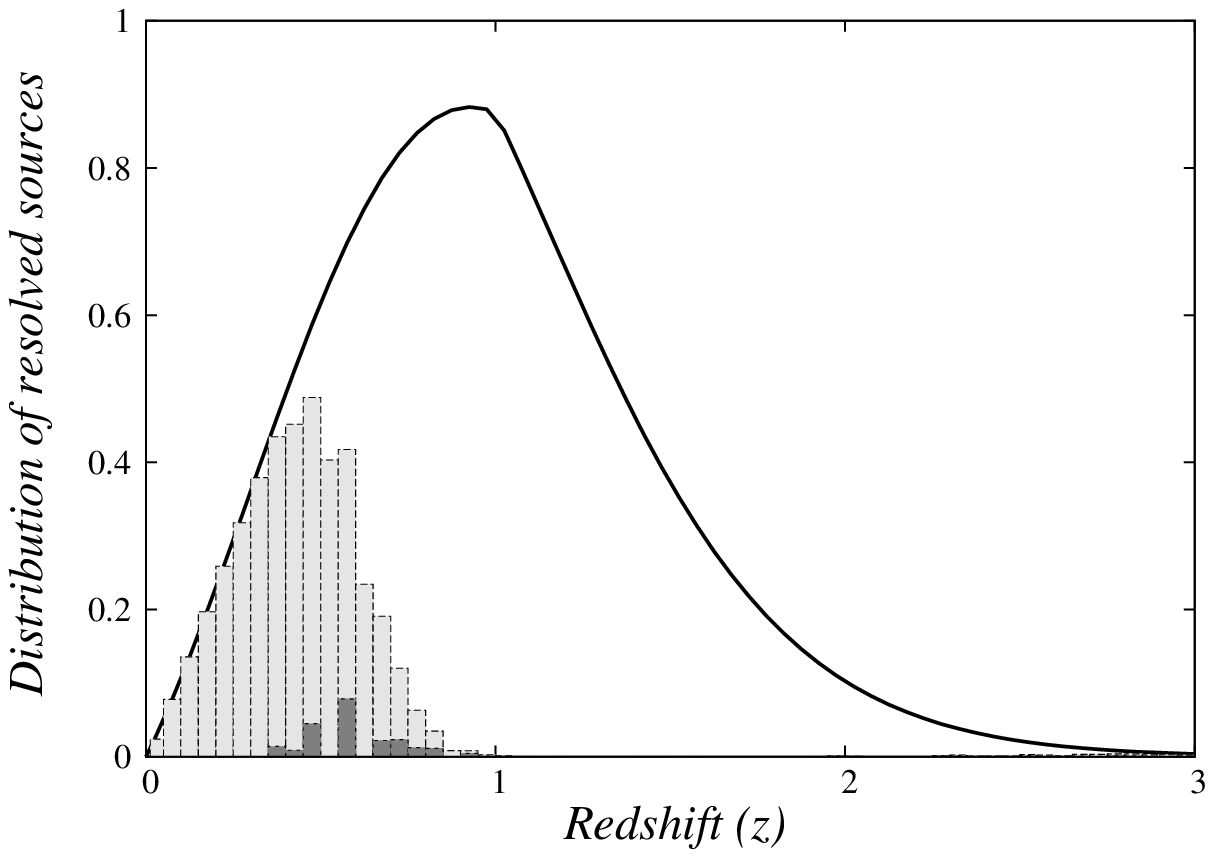}%
\includegraphics[width=0.50\textwidth]{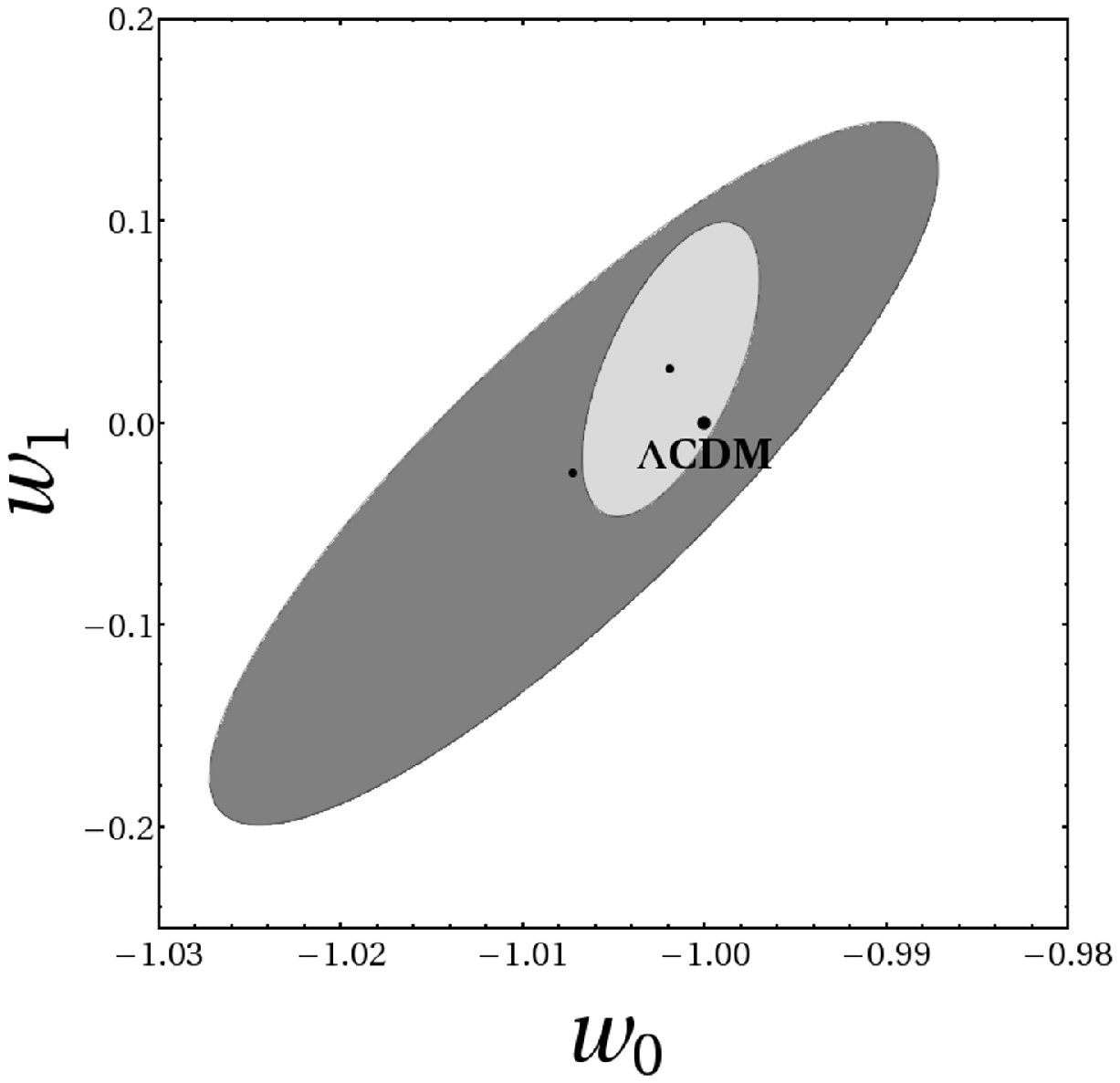}%
\caption{{\it Left panel} displays the distribution of resolved sources
(light grey) and misidentified sources (dark grey) for $n=10$ in
(\ref{eq:BBO}) and $m=3$ in (\ref{eq:redshift_range}). (Note the lack of
resolved sources beyond $z=1$.) The {\it right panel} shows confidence
contours for the EOS of dark energy with $n=10$ and $m=1$ (light grey)
and $m=3$ (dark grey), respectively.}
\label{fig:10BBO3}
\end{figure*}

In Fig~\ref{fig:7.eps} we plot the constraints obtained by choosing
different beams sizes but keeping the redshift error box at one sigma
corresponding to $m=1$ in (\ref{eq:redshift_range}). The beam size
chosen for different one sigma regions is $n=1,2,3,10$ in
(\ref{eq:BBO}), and we find that the area under the one sigma confidence
level increases with $n$. This is due to the fact that the total number
of GPs decreases with increasing beam size. For this figure the
misidentified sources were included in the determination of cosmological
parameters.  We find that even for $n=10$, which is large enough to
ensure that the true source would almost certainly fall within the beam,
the constraints are sufficiently narrow, with very little bias, even
though as Fig~\ref{fig:10BBO} illustrates, the total number of resolved
sources in this case is smaller than what one finds for $n=1,2,3$
(compare with figure \ref{fig:RES_BAD}); and the number of
misidentifications is large, due to the fact that we have chosen $m=1$
in (\ref{eq:redshift_range}).

The left panel of Fig~\ref{fig:10BBO3} plots the distribution of
resolved sources for $n=10$ and $m=3$, corresponding to the largest
error box that we have considered. The number of resolved sources has
decreased substantially, however, on the positive side, the number of
misidentifications is much smaller than in Fig~\ref{fig:10BBO}. Note
also that there are almost no resolved sources beyond $z=1$. The
reason for this is the large source-redshift error at high redshift,
which, when taken together with the fact that we are allowing the source
to fall in the three sigma redshift range, makes it very difficult to
resolve sources. The right panel of the same figure shows the
constraints obtained on cosmology. As expected, the constraints are
poorer due to lack of sources beyond $z=1$.

The main reason for choosing large $n$ is to ensure that the each beam
contains the true source at a greater probability. Our results show that
for $n=10$ and $m=1$ the number of resolved sources is sufficient to give
good constraints on cosmology, however, the number of misidentifications
is large. The bias thus produced is not large and it would seem that
this configuration may give us good unbiased constraints on cosmology.
However, since our fiducial model has only two parameters, it is
possible that the bias may make determining cosmology more difficult
for more complicated dark energy models. Therefore, it is necessary to
keep the misidentifications small. It is difficult to make general
statements regarding the optimum configuration for $n$ and $m$ for a
more complicated dark energy model, however, our results suggest that
the range $ 3 \le n  \le 6$ and $ 2 \le m \le 3$ may work in most cases.

\section{Conclusions}

To conclude, we would like to underscore the important point that our
self-calibrating scheme works very well {\em even if none} of the
gravity wave sources have observable electromagnetic signatures. Indeed
if the beam width is not too large ($\leq$ ten times BBO), then the
presence of only a few gold plated sources (those whose redshift has
been independently established) in conjunction with the iterative
procedure presented here, allows us to determine the equation of state
of dark energy to an accuracy of a few percent -- see figure
\ref{fig:7.eps}. The two main issues investigated in this work are the
application of ideas presented in S3 to simulated data, and to address
the issue of misidentification of GW sources. Our simulations simplify
details in two respects: first, we do not include the effect of
clustering, second, we model the lensing scatter as a symmetric Gaussian
distribution. (We shall assess the effects of asymmetry induced my
lensing as well as the effects of clustering in a companion paper.) We
find that the method works quite well on simulated data and the
iterations saturate after a small number of steps.

We showed that misidentified sources might in general bias the
determination of DZ relation and would lead to a runaway process by
which the DZ relation would move away systematically from the true
cosmological DZ relation, thus biasing the estimation of the
cosmological parameters. In our simulations, the effect of biasing is
present but is found to be small. We have shown that increasing the
allowed redshift range for the GW source reduces this
bias even further, without significantly affecting the cosmological
constraints (due to a reduction in the number of resolved sources). A
further positive conclusion is that the method works even for larger
beam sizes, thereby addressing the concern that not all beams  might
contain the source galaxy.

In our simulations the asymmetric lensing scatter has been modeled as
a symmetric Gaussian distribution. Lensing scatter has a large number
of demagnified sources and a small number of compensating large
magnifications. As mentioned in \cite{holz09}, highly magnified
sources would show up as outliers in the DZ plot and can be either
removed from the sample or else, in the method that we propose, can be
handled by choosing a small value of $m$. As our results show, the
bias due to misidentifications is relatively small, so this choice
would not cause significant distortions in the estimates of
cosmological parameters.

\section{Acknowledgment}
Tarun D. Saini is thankful to the Associateship Programme of IUCAA for
providing support for his stay at IUCAA where part of this work was
done.


\begin{thebibliography}{}
\bibitem{DE}
V. Sahni and A.~A.~Starobinsky, Int. J. Mod. Phys. D {\bf 9}, 373
(2000);
P.~J~E.~Peebles and B.~Ratra, Rev. Mod. Phys. {\bf 75}, 559 (2003);
T. Padmanabhan, Phys. Rep. {\bf 380}, 235 (2003);
V. Sahni, Lect. Notes Phys. {\bf 653}, 141 (2004);
V. Sahni,  astro-ph/0502032;
E.~J.~Copeland, M.~Sami and S.~Tsujikawa, Int. J. Mod. Phys. D {\bf 15}, 1753 (2006);
J.~A.~Frieman, M.~S.~Turner and D.~Huterer, Ann. Rev. Astron. Astroph.
{\bf 46}, 385 (2008);
R. Durrer and R. Maartens,
"Dark Energy: Observational and Theoretical Approaches", ed. P Ruiz-Lapuente (Cambridge UP, 2010), pp. 48 - 91
[{\tt arXiv:0811.4132}];
S. Nojiri and S. D. Odintsov, Phys.Rept. {\bf 505}, 59 (2011) [e-Print: arXiv:1011.0544];
T. Clifton, P. G. Ferreira, A. Padilla and C. Skordis,
Phys.Rept. {\bf 513} 1 (2012) [e-Print: arXiv:1106.2476].
\bibitem{DE1} V.~Sahni and A.~A.~Starobinsky, Int. J. Mod. Phys. D {\bf
15}, 2105 (2006).
\bibitem{schutz86} B. F. Schutz, Nature, 323, 310 (1986)
\bibitem{decigo} N. Seto, S. Kawamura, and T. Nakamura, Phys. Rev. Lett,
87 221103 (2001); S. Kawamura, et al.,
Journal of Physics: Conference Series 122, 012006 (2008);
R. Takahashi and T. Nakamura, Prog. Theor. Phys. {\bf 113}, 63 (2005).
\bibitem{holz051} D. E. Holz and S. A.  Hughes, \apj,  629, 15 (2005)
\bibitem{arun}K. Arun, B. Iyer, B. Sathyaprakash, S, Sinha and C. Van Den Broeck,
Phys. Rev	. D 76, 104016 (2007); Erratum-ibid.D76:129903,2007
\bibitem{sathya} B.S. Sathyaprakash, B.F. Schutz and C. Van Den Broeck,
Class.Quant.Grav. {\bf 27}, 215006, (2010).
\bibitem{lisa}T. Apostolatos, C. Cutler, G. Sussman and K. Thorne,
Phys. Rev. D {\bf 49}, 6274 (1994)
\bibitem{elisa} P. Amaro-Seoane, et al., Class. Quantum Grav. {\bf 29}, 124016 (2012).
\bibitem{BBO} E.S. Phinney et~al., 2003, {\em Big Bang Oberver Mission Concept
Study} (NASA); 
A. Nishizawa, K. Yagi, A. Taruya and T. Tanaka, Phys. Rev. D {\bf 85}, 044047 (2012);
A. Nishizawa, A. Taruya and S. Saito, Phys. Rev. D {\bf 83}, 084045 (2011);
J. Crowder and N.J. Cornish, Phys. Rev. D {\bf 72}, 083005 (2005).
\bibitem{Ni} Wei-Tou Ni, arXiv:1104.5049.
\bibitem{cutler2007}
C. Cutler;M.  Vallisneri, Phys. Rev. D., 76, 104018 (2007)
\bibitem{holz09} C. Cutler, D.  E. Holz\ 2009, arXiv:0906.3752v1
[astro-ph.CO]
\bibitem{hirata2010} C. M. Hirata; D.E. Holz; C. Cutler, Phys. Rev. D.,
81, 124046 (2010)
\bibitem{zhao11} Zhao, W., van den Broeck, C., Baskaran, D., \& Li,
T.~G.~F., Phys. Rev. D., 83, 023005 (2011)
\bibitem{bense08} B. Kocsis, Z.  Haiman, and K.  Menou, \apj, 684, 870 (2008)
\bibitem{macleod} C.L. MacLeod and C.L. Hogan, Phys. Rev. D., 77, 043512 (2008)
\bibitem{saini10} T.D. Saini, S.K. Sethi, \& V. Sahni, Phys. Rev. D., 81, 103009 (2010) 
\bibitem{holz05} D.E. Holz, and E.V. Linder, \apj, 631, 678 (2005)
\bibitem{CPL}
M. Chevallier and D. Polarski, 2001, Int. J. Mod. Phys. D {\bf 10}, 213
[{\tt gr-qc/0009008}];
E. V. Linder, 2003, \prl {\bf 90}, 091301
[{\tt astro-ph/0208512}].
\bibitem{kaiser} N. Kaiser, ApJ, 388, 272, (1992)
\bibitem{hu} W. Hu, ApJ, 522, L21, (1999)
\bibitem{hudf} S.V.W. Beckwith et al. AJ, 132, 1729, (2006).
\bibitem{mil81} K.S. Miller, Mathematics Magazine, Vol. 54, No. 2, pp. 67-72
\end{thebibliography}
\end{document}